\documentclass[a4paper,10pt,twocolumn]{article}

\usepackage[english]{babel}
\usepackage[utf8]{inputenc}
\usepackage[T1]{fontenc}

\usepackage[top=1.5cm, left=1.5cm, right=1.5cm, bottom=1.5cm]{geometry}

\renewenvironment{abstract}{\bf\small {\em\ Abstract---}}{}

\usepackage{amsfonts,amssymb,amsmath,amsthm}
\usepackage{subfigure}
\usepackage{graphicx}
\usepackage[footnotesize]{caption}

\usepackage{amsfonts,amssymb,amsmath,amsthm}

\usepackage{hyperref} 
\usepackage{dsfont} 
\usepackage{amssymb} 
\usepackage{bm} 
\usepackage{bbm} 
\usepackage{algorithm}
\usepackage{algpseudocode}
\usepackage{mathtools}
\usepackage{color}

\def\x{{\mathbf x}}
\def\z{{\mathbf z}}
\def\y{\mathbf{y}}

\def\u{\mathbf{u}}

\def\D{{\mathbf D}}

\def\I{{\mathbf I}}

\def\R{{\mathbb R}}

\DeclareMathOperator{\Mr}{\mathbf{M}^{\textrm{r}}}
\DeclareMathOperator{\Mcp}{\mathbf{M}^{\textrm{u}}}
\DeclareMathOperator{\Mcm}{\mathbf{M}^{\textrm{l}}}
\DeclareMathOperator{\thetap}{\theta^{\textrm{u}}}
\DeclareMathOperator{\thetam}{\theta^{\textrm{l}}}

\DeclareMathOperator{\bfalpha}{\bm{\alpha}}

\DeclareMathOperator*{\argmin}{argmin}
\DeclareMathOperator*{\sign}{sign}
\DeclareMathOperator*{\dEU}{d}

\title{Fast Iterative Shrinkage for Signal Declipping and Dequantization}

\author{Lucas Rencker$^1$\thanks{The research leading to these results has received funding from the European Union's H2020 Framework Programme (H2020-MSCA-ITN-2014) under grant agreement no 642685 MacSeNet.}, Francis Bach$^2$, Wenwu Wang$^{1}$, and Mark D. Plumbley$^1$.\\
  \footnotesize $^{1}$Centre for Vision, Speech and Signal Processing, University of Surrey, Guildford, UK\\ \footnotesize$^2$SIERRA-project team, INRIA, Paris, France} \date{\empty} 

\begin{document}

\maketitle

\begin{abstract} 
We address the problem of recovering a sparse signal from clipped or quantized measurements. We show how these two problems can be formulated as minimizing the distance to a convex feasibility set, which provides a convex and differentiable cost function. We then propose a fast iterative shrinkage/thresholding algorithm that minimizes the proposed cost, which provides a fast and efficient algorithm to recover sparse signals from clipped and quantized measurements.
\end{abstract}

\section{Introduction}
\label{sec:introduction}

Clipping and quantization are common distortions in digital signal processing. In this paper we address the problem of recovering a signal from clipped or quantized measurements. We consider a distorted signal $\y = f(\x)$ where $\x \in \R^N$ is the original clean signal and $f: \R^N \mapsto \R^N$ is a nonlinear and non-invertible clipping or quantization function. We further assume that $\x$ is sparse with respect to a known \emph{overcomplete} dictionary $\D \in \R^{N \times M}$ ($N<M$), i.e., $\x = \D \bfalpha$ with $\bfalpha$ sparse.

Recovering a sparse signal from clipped or quantized measurements is often formulated as the following constrained sparse coding problem \cite{2015_Kitic_Sparsity,2016_Moshtaghpour_Consistent}:
\begin{equation}\label{eq:constrained}
\min_{\bfalpha} \Psi(\bfalpha) \quad \text{s.t.} \quad \D\bfalpha \in f^{-1}(\y)
\end{equation} 
where $\Psi(\cdot)$ is a sparsity-inducing norm or pseudo-norm, and $f^{-1}(\y)$ is the \emph{pre-image} of the observed signal $\y$ through $f$. The set $f^{-1}(\y)$ can be seen as the \emph{feasibility set} associated with the measurement $\y$, i.e., the set of possible input signals that could have generated $\y$. In the case of declipping, the feasibility set can be explicitly formulated as $f^{-1}(\y) = \{\x|\Mr \x = \Mr \y, \Mcp \x \succeq \thetap \mathbf{1}, \Mcm \x \preceq \thetam \mathbf{1}\}$ \cite{2015_Kitic_Sparsity} where $\Mr$, $\Mcp$ and $\Mcm$ are diagonal binary matrices indicating the reliable, upper and lower clipped samples respectively, and $\thetap>\thetam$ are upper and lower clipping thresholds respectively. In the case of quantization, $f^{-1}(\y) = [l_1, u_1)\times...\times[l_N, u_N)$, where $[l_i, u_i)$ is the quantization region associated with each sample $y_i$.

Eqn. \eqref{eq:constrained} is a constrained, non-smooth and possibly non-convex optimization problem which can be difficult to solve. Recently, algorithms based on Alternating Direction Method of Multipliers (ADMM) \cite{2011_Boyd_Distributed} or the related Douglas-Rachford algorithm \cite{2007_Combettes_Douglas} have been proposed to solve \eqref{eq:constrained}, see \cite{2015_Kitic_Sparsity} for declipping or the implementation of \cite{2016_Moshtaghpour_Consistent} in \cite{2014_Perraudin_UNLocBoX} for dequantization. However, these algorithms involve computing proximal operators of the type
\begin{equation}\label{eq:projection}
\argmin_{\bfalpha} \|\u - \bfalpha\|_2^2 + \mathbbm{1}_{f^{-1}(\y)}(\D\bfalpha)
\end{equation} 
at each iteration for some $\u \in \R^M$ (see e.g. \cite{2015_Kitic_Sparsity}), where $\mathbbm{1}_{f^{-1}(\y)}(\cdot)$ is the indicator function of the set $f^{-1}(\y)$. When the dictionary $\D$ is orthogonal or a tight frame ($\D^T\D = \I$), \eqref{eq:projection} can be computed efficiently in closed form. However for general overcomplete dictionaries, \eqref{eq:projection} is a non-orthogonal projection which has to be computed iteratively, using (e.g.) another nested ADMM algorithm at each iteration. This leads to a heavy computational cost, which can be prohibitive for large-scale applications. 

\section{Proposed problem formulation}
\label{sec:problem_formulation}

We propose to relax the constrained problem \eqref{eq:constrained} as the following unconstrained problem (already proposed in \cite{2018_Rencker_Consistent} in the context of declipping):
\begin{equation}\label{eq:problem}
\min_{\bfalpha} \frac{1}{2} \dEU(\D\bfalpha,f^{-1}(\y))^2 + \lambda \Psi(\bfalpha)
\end{equation}
where $\dEU(\x,\mathcal{C})^2$ is the squared Euclidean distance between $\x$ and the set $\mathcal{C}$, defined as:
\begin{equation}\label{eq:distance}
\dEU (\x, \mathcal{C})^2 = \min_{\z \in \mathcal{C}} \|\x-\z\|^2_2.
\end{equation} 
The proposed formulation thus enforces the estimated signal $\D\bfalpha$ to be close to its feasibility set $f^{-1}(\y)$, where the parameter $\lambda$ controls a trade-off between data fidelity and sparsity. Note that when $\lambda \rightarrow 0^+$, \eqref{eq:problem} is equivalent to the constrained problem \eqref{eq:constrained}. However since $f^{-1}(\y)$ is convex, the proposed problem formulation provides convenient properties which we recall here:
\begin{itemize}
\item The data-fidelity term $\dEU (\x, f^{-1}(\y))^2$ is convex, as a minimum of a family of convex functions $\|\cdot\|_2^2$ over a non-empty and convex set \cite[Section 3.2.5]{2004_Boyd_Convex}
\item $\dEU (\cdot, f^{-1}(\y))^2$ is differentiable with gradient \cite[eqn. (1.1)]{1973_Holmes_Smoothness}:
\begin{equation}\label{eq:gradient}
\nabla_{\x} \frac{1}{2} \dEU (\x, f^{-1}(\y))^2 = \x - \Pi_{f^{-1}(\y)}(\x),
\end{equation}
where $\Pi_{f^{-1}(\y)}(\x)$ is the orthogonal projection of $\x$ onto $f^{-1}(\y)$.
\item The gradient \eqref{eq:gradient} is 1-Lipschitz. This stems from the contraction property of projection operators onto convex sets, see e.g. \cite[Prop. B.11]{1999_Bertsekas_Nonlinear}.
\end{itemize}
The proposed formulation is thus a problem of minimizing a convex and differentiable data-fidelity term, along with a sparsity-inducing regularizer, which is similar to classical sparse recovery methods such as Basis Pursuit Denoising (BPDN) \cite{2001_Chen_Atomic}. Moreover when the feasibility set is a singleton $f^{-1}(\y) = \{\x\}$ (i.e., the signal is unclipped/unquantized), then \eqref{eq:problem} simplifies to a classical sparse recovery problem such as BPDN:
\begin{equation}\label{eq:BPDN}
\min_{\bfalpha} \frac{1}{2} \|\D\bfalpha-\x\|_2^2 + \lambda \Psi(\bfalpha).
\end{equation}

\section{Proposed algorithm}
\label{sec:alg}

In the rest of this paper we focus on the convex $\ell_1$ case, i.e. solving \eqref{eq:problem} with $\Psi(\bfalpha) = \|\bfalpha\|_1$. In this case \eqref{eq:problem} becomes a problem of minimizing the sum of a convex and smooth cost data-fidelity term, along with a convex and non-smooth regularizer. This can be classically solved using Iterative Shrinkage/Thresholding Algorithms (ISTA) \cite{2009_Beck_fast}. ISTA is an attractive class of algorithm since they only involve gradient computations and simple element-wise thresholding, making them simple and adequate for large-scale problems. More precisely, ISTA applied to \eqref{eq:problem} iterates (after an initial guess $\bfalpha_0$):
\begin{equation}\label{eq:ISTA}
\bfalpha_{k+1} = S_{\mu\lambda}\big(\bfalpha_{k}-\mu \D^T(\D\bfalpha_k - \Pi_{f^{-1}(\y)}(\D\bfalpha_k))\big)
\end{equation}
where $S_{\rho}(.)$ is the soft-thresholding operator:
\begin{equation}
S_{\rho}(\bfalpha)_i = \max(|\alpha_i|-\rho,0)\sign(\alpha_i),
\end{equation}
and $\mu$ is a step size that can be typically set as $1/L$, where $L = \|\D^T\D\|_2$ is the Lipschitz constant of the gradient \cite{2009_Beck_fast}. Note that $\Pi_{f^{-1}(\y)}(\cdot)$ here is a simple orthogonal projection that can be computed using element-wise maxima. The ISTA type algorithm \eqref{eq:ISTA} thus provides a simple and efficient way to solve the relaxed problem \eqref{eq:problem}, which is computationally simpler than ADMM based algorithms to solve the constrained problem \eqref{eq:constrained}. However, for badly conditioned matrices $\D^T\D$, ISTA is also known to converge quite slowly. Several algorithms have been proposed to speed up ISTA, such as the celebrated \emph{Fast Iterative/Shrinkage Thresholding} algorithm (FISTA) \cite{2009_Beck_fast}. FISTA applied to the proposed problem is presented in Algorithm \ref{alg:FISTA}.
\begin{algorithm}[H]
	\caption{FISTA for declipping/dequantization}\label{alg:FISTA}
	\begin{algorithmic}
		\Require $f^{-1}(\y), \D, \lambda, \bfalpha_0$
		\State \textbf{Initialize:} $\u_1 = \bfalpha_0, t_1 = 1, k = 1$
		\State \textbf{Iterate} until convergence:
		\begin{align}
		&\bfalpha_{k} =  S_{\mu\lambda}\big(\u_{k}-\mu \D^T(\D\u_k - \Pi_{f^{-1}(\y)}(\D\u_k))\big)\label{eq:FISTA_1}\\
		&t_{k+1} = \frac{1+\sqrt{1+4t_k^2}}{2}\label{eq:FISTA_2}\\
		&\u_{k+1} =  \bfalpha_k + \bigg(\frac{t_k-1}{t_{k+1}}\bigg)(\bfalpha_k-\bfalpha_{k-1})	\label{eq:FISTA_3}\\
		&k =  k + 1	
		\end{align}
		\State \textbf{return} $\hat{\bfalpha}$
	\end{algorithmic}
\end{algorithm}
FISTA thus simply computes the thresholded gradient descent step on a linear combination of the two previous estimates $\bfalpha_k$ and $\bfalpha_{k-1}$. Note that the cost of computing \eqref{eq:FISTA_2} and \eqref{eq:FISTA_3} is negligible compared to that of \eqref{eq:FISTA_1}, so FISTA does not incur extra computational cost per iteration compared to ISTA. However, the convergence rate of FISTA can be shown to be of $\mathcal{O}(1/k^2)$, instead of $\mathcal{O}(1/k)$ for ISTA \cite{2009_Beck_fast}.

\section{Numerical results}

\begin{figure}[h!]
	\centering
	\includegraphics[width=0.45\textwidth]{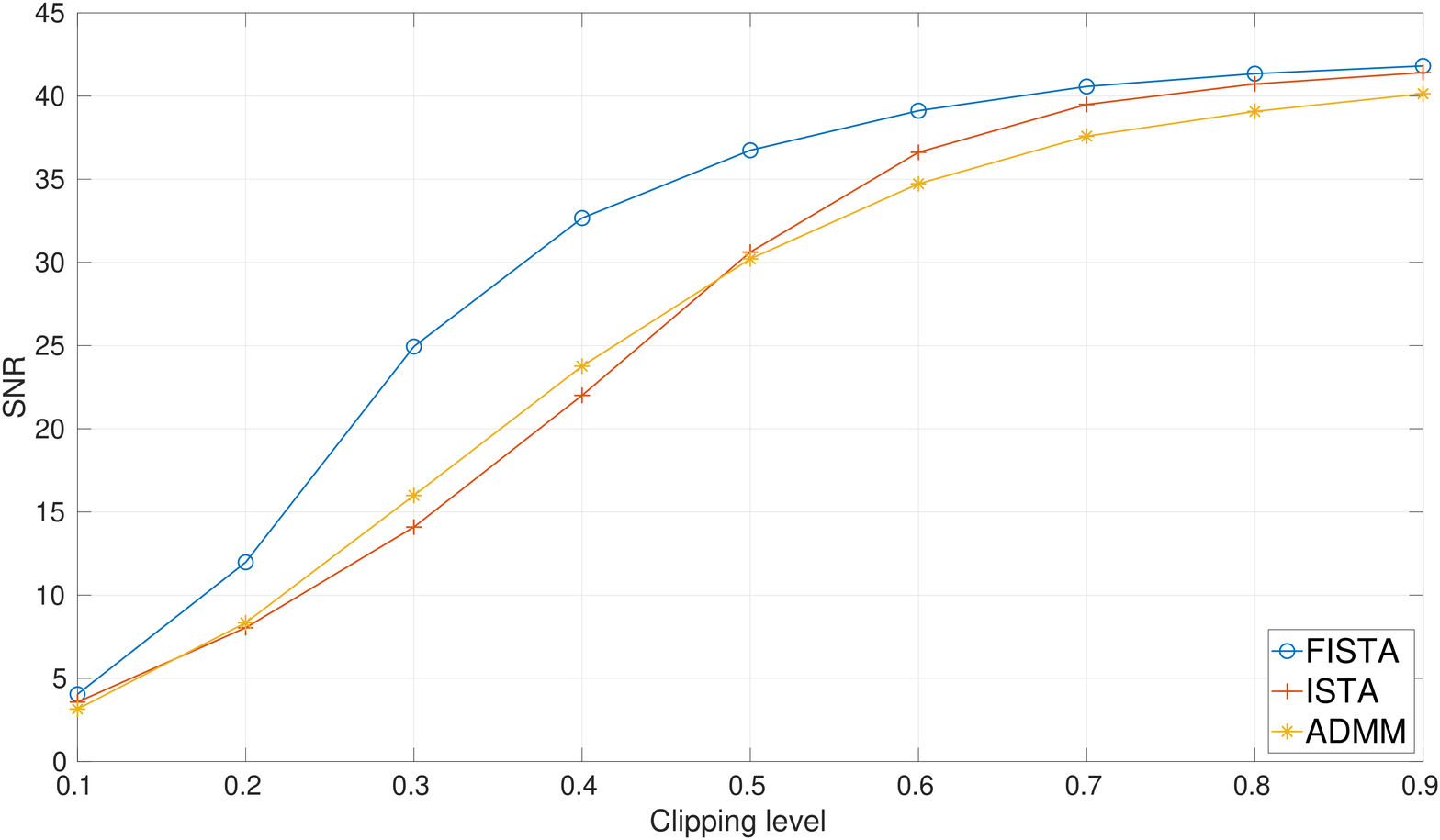}
	\caption{Declipping performance for different clipping levels}
	\label{fig:declipping}
\end{figure}

\begin{figure}[h!]
	\centering
	\includegraphics[width=0.45\textwidth]{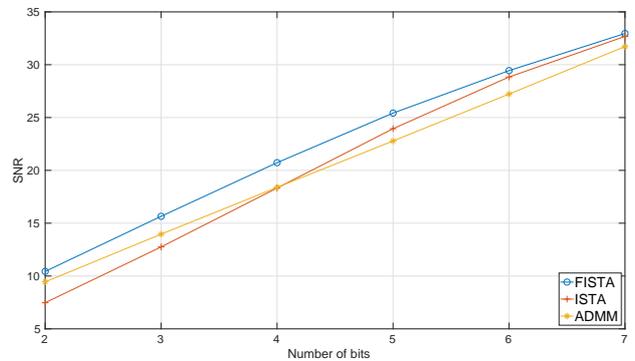}
	\caption{Dequantization performance for different quantization levels}
	\label{fig:dequantization}
\end{figure}

We generate a random dictionary $\D \in \R^{256 \times 512}$ with Gaussian i.i.d entries and 100 16-sparse vectors $\bfalpha \in \R^{512}$, and normalize the resulting signals $\x = \D\bfalpha$ to unit $\ell_{\infty}$ norm. We then clip or quantize each vector as $\y = f(\x)$, using clipping at different clipping level $\theta$, and a uniform midriser quantizer with bin width $\Delta = 2^{1-N_b}$, where $N_b$ is the number of bits. We compare the constrained formulation \eqref{eq:constrained} solved using ADMM (see e.g. \cite{2015_Kitic_Sparsity}), and the proposed relaxed approach \eqref{eq:problem} solved using (F)ISTA. All algorithms are computed using the $\ell_1$-norm, and we fix $\lambda = 10^{-2}$ for (F)ISTA. The algorithms are evaluated in terms of average SNR of the reconstructed signals $\hat{\x} = \D\hat{\bfalpha}$. The results are presented in Figure \ref{fig:declipping} and \ref{fig:dequantization}.


All algorithms are computed with a maximum of 400 iterations, or until the algorithm has converged. Experiments show that the proposed formulation \eqref{eq:problem} solved with ISTA leads to comparable results to ADMM. However due to its slow convergence rate, ISTA might still be far from the optimum even after 400 iterations. FISTA on the other hand often reaches the optimum in under 150 iterations, and leads to a significant performance increase in terms of signal reconstruction. The average computational time for each algorithm is reported in Table \ref{table}. As expected, solving \eqref{eq:problem} using (F)ISTA is significantly faster than solving \eqref{eq:constrained} using ADMM, since (F)ISTA only involves computing gradients and element-wise computations, while ADMM involves computing non-orthogonal projections at each iteration.

\begin{table}[h!]
\centering
\begin{tabular}{|c||c|c|c|}
\hline
cpu time (s) & ADMM & ISTA & FISTA\\
\hline\hline
declipping & 968.4 & 3.79 & \bf{1.56}\\
\hline
dequantization & 943.52 & 3.88 & \bf{1.81}\\
\hline
\end{tabular}
\caption{Average computational time of each algorithm}
\label{table}
\end{table}

\section{Conclusion}
\label{sec:conclusion}

We showed that relaxing the constrained problem \eqref{eq:constrained} leads to a simple optimization problem, which can be solved using  a fast iterative shrinkage algorithm. The proposed algorithm leads to increased performance and a significant reduction in computational time.

\bibliographystyle{IEEEtran}
\bibliography{IEEEabrv,../../References/bibliography.bib}

\end{document}